\documentclass[prl,twocolumn,amsmath,amssymb,aps]{revtex4}
\usepackage{graphicx}

\begin{document}

\title{Dawn of Cavity Spintronics}

\author{C.-M. Hu$^{\footnote{Electronic address: hu@physics.umanitoba.ca;
URL: http://www.physics.umanitoba.ca/$\sim$hu}}$}

\affiliation{Department of Physics and Astronomy, University
of Manitoba, Winnipeg, Canada R3T 2N2}

\date{\today}

\begin{abstract}
Merging the progress of spintronics with the advancement in cavity quantum electrodynamics and cavity polaritons, a new field of Cavity Spintronics is forming, which connects some of the most exciting modern physics, such as quantum information and quantum optics, with one of the oldest science on the earth, the magnetism.
\end{abstract}

\maketitle

\section{Heroes of Spintronics }

Heroes are not always stars. This we know from movie, but it is also true in the community of physicists. While stars are celebrated and attract fans, heroes are more personal, reflecting our own views on the meaning of life and value of work. We don't know each other's heroes, and their path may not be easy to follow. Some of them may even prefer to stay away from the spotlight. My heroes, in the field of spintronics, are Robert Silsbee and Mark Johnson.

In 1979, Robert H. Silsbee, the 50-year old Cornell University professor on his sabbatical leave at the University of Paris-Sud, published an article in Physics Review B with the title \textit{Coupling between ferromagnetic and conduction-spin-resonance modes at a ferromagnetic-normal-metal interface} \cite{Silsbee1979}. Using a clever experiment and a concise theoretical model, Silsbee et al. revealed two new physical characteristics of a spin current: (i) through the microwave excitation of ferromagnetic resonance (FMR), a spin current can be generated in a ferromagnetic metal, which will flow into an adjacent normal metal. (ii) Such a spin current pumped by FMR impacts the spin dynamics of electrons in the metal via the exchange interaction, which can be detected via an enhanced electron spin resonance. Several decades later, from these two effects came \textit{spin pumping} \cite{Tserkovnyak2002} and \textit{spin torque} \cite{Berger1996}, two seminal concepts of contemporary magnetism. Today, every conference on spintronics has celebrated talks of spin pumping. However, Silsbee \textit{et al}.'s ground breaking paper has only been cited about 100 times in four decades. Not surprising, many students of today studying spin pumping have never read that paper, and some have never even heard of it. The same thing has happened in the history of cinematography. In an era when movie goers celebrate blockbusters of Steven Spielberg, George Lucas and Peter Jackson, few of us spend time at home enjoying Stanley Kubric's \textit{2001, the Space Odyssey}, a ground breaking art work produced in 1968, which has inspired Spielberg and Lucas.

Kubric directed only 16 movies in his life time. He was too meticulous to endure anything but a masterpiece. So was Silsbee, who published only 82 papers. Silsbee never cared about the number of his papers, he just loved doing original work. His 1979 paper on spin pumping \cite{Silsbee1979} demonstrated optical injection and detection of spin current. That work inspired Silsbee's vision of electrical spin injection and detection, in which a spin current would be generated by a battery and detected by a voltmeter, without using either microwaves or light. 20 years later, that vision became the main topic of spintronics \cite{Wolf2001}. But by then, it was a bold idea. The giant magnetoresistance effect (awarded the 2007 Nobel Prize) would be discovered nine years later \cite{Baibich1988}, and the word \textit{spintronics} did not exist. So in 1980, when Silsbee led his PhD student Mark Johnson to explore electrical spin injection into metals, it was not a race. It was a courageous lonely journey into a no man's land, as research in physics is meant to be.

Johnson's adventure in Ithaca working on spin injection was a scientific odyssey reminiscent of the epic of Homeric heroes. His achievement came after multiple years of failure. When I study his thesis work published in 1985 in Physics Review Letters (PRL) entitled \textit{Interfacial Charge-Spin Coupling: Injection and Detection of Spin Magnetization in Metals} \cite{Johnson1985}, I realize that truly original work demands courage more than ideas. If I put myself in Silsbee's shoes as a professor in 1980, I might have had the same vision, but I would not have had the courage to realize it.

The best way to appreciate Johnson and Silsbee's originality, creativity and especially their courage is to read their 1985 PRL paper \cite{Johnson1985} in comparison with a paper published in 2002 by Jedema \textit{et al}. in Nature \cite{Jedema2002}. In the history of cinematography, classics are sometimes reproduced to make blockbusters using new technologies. The same is true in science. Senior movie fans of \textit{King Kong} may remember Jessica Lange's marvellous performance in 1976, in her first film cast in the adventure facing a beast (and if you are really old, you may even remember Fay Wray). At that time, the amazing digital visual effects used in 2005 by Peter Jackson were not yet available. This mirrors the time of Johnson's thesis work. At that time, Silsbee predicted that the spin accumulation was inversely proportional to sample volume, but in 1980 there was no clean room at Cornell University for making nano-structured devices, so Johnson developed his own lithography for making small samples. The next challenge for him was in the measurement. Even in the smallest sample he made for spin injection (where all three dimensions are the order of 100 $\mu$m), the signal was only a few picovolts. To measure it, Johnson designed a special bridge circuit, and used it together with a SQUID voltmeter and a lockin to beat down noise. The third big obstacle was the spurious magnetoresistance effect in his device, which hampered his progress for years until he invented the non-local detection method (together with Andras Janossy). Finally, to conclusively verify the spin signal, he and Silsbee introduced Hanle effect \cite{Hanle1924} from atomic physics into condensed matter physics. Such a spin precession effect was the golden compass that guided their courageous adventure in spin transport \cite{Johnson1985}.

Decades later, nonlocal detection combined with Hanle effect would become the norm of performing spin injection and detection experiments. This was neatly shown in Nature's reproduction in 2002 \cite{Jedema2002}. Comparing this exquisite work \cite{Jedema2002} with the original paper of Johnson and Silsbee \cite{Johnson1985}, it is clear that spintronics is a nano science as Silsbee has predicted. By shrinking the device size from micrometers to nanometers, Johnson's picovolt spin signal detected in 1985 at cryogenic condition was enhanced by four orders of magnitude, so that in 2002 it was measured at room temperature. Today, with the advancement in nanotechnology, spin current is not only routinely measured in the labs by students, but it is becoming an integral part of the electronics that we use daily \cite{Bader2010}. In nano-structured magnetic devices, spintronic effects are often so large, so important, and so useful, that they are transforming the old science of magnetism, and thereby innovating our information and communication technology \cite{Bader2010}.

That is great. The only downside is, as any research field becomes crowded, new discoveries are increasingly achieved in hectic races, so that meticulous work is often over shadowed by rushed publishing, and the spirit of science gives way to sleazy business of number counting. As in George Lucas' expanded universe (also known as the EU) of Star Wars, once the legions of clone army become the main force in conquering new worlds, the Jedi Master and his Predawn leave the stage. So in the rapidly expanding kingdom of spintronics, the heroic spirit of Silsbee and Johnson is often forgotten, or ignored.

But as in Lucas' EU, there comes a new hope. In the remote area where the galaxy of spintronics boarders the alien territories of different physics, the boots of the clone army have not yet arrived. In that quiet space, plenty of mysteries are waiting to be explored by the free minds of science. It is there that the twilight of Cavity Spintronics is being seen, where the progress of Spintronics is merging with the advancement in two different fields: Cavity Quantum Electrodynamics (QED) and Cavity Polaritons.

\section{Stars and Innovators in Cavity Quantum Electrodynamics}

Lacking personal interactions with the great physicists of Cavity QED, my knowledge is limited to the work of its well-known stars. In 2012, the Royal Swedish Academy of Science decided to award the Nobel Prize in Physics to Serge Haroche and David Wineland. They both use traps to study particles in quantum states. Wineland electrically traps charged atoms, controlling and measuring them with photons \cite{Wineland2013}. Haroche takes the opposite approach; he controls and measures trapped photons by sending atoms through a microwave cavity \cite{Haroche2013}. Their ingenious trap and cavity techniques allow the fundamental interaction between light and matter to be studied in its most elementary form in the quantum regime. Reaching this regime has generated the field of cavity QED \cite{Cohen-Tannoudji2004}, and opened the door to a new era of using coherent quantum effects for quantum information processing \cite{Laflamme2002}.

As Johnson and Silsbee's journey has shown, innovative condensed matter physicists are often inspired by atomic effects. Eight years before Haroche received the Nobel Prize, his work had inspired Rob Schoelkopf at Yale University. In a paper published in 2004 in Nature entitled \textit{Strong coupling of a single photon to a superconducting qubit using circuit quantum electrodynamics} \cite{Wallraff2004}, Schoelkopf and his team show how to take cavity QED from the atomic world to a solid-state system. They used a superconducting two-level system to play the role of an artificial atom, and studied how it was quantum coherently coupled to a single microwave photon confined in a cavity made of a superconducting transmission line resonator. Such an innovative on-chip technique paved the way to circuit QED, and it opened many new possibilities for studying the strong interaction of light and matter in a variety of solid-state systems. Reading Schoelkopf's paper, some of us working on spintronics immediately had the idea of using a spin two-level system in magnetic materials to couple with photons in the microwave cavity. Doing so would propel the research in magnetization dynamics into a completely new regime of quantum coherent spin-photon coupling, merging spintronics with circuit QED to advance both fields. In 2010, before any experimental pioneers set their footprints on the new land, theoreticians prepared a map of quantum physics \cite{Soykal2010} for the dream world of Cavity Spintronics.

\section{Settlers of the Cavity Spintronics}

In their theoretical paper entitled \textit{Strong Field Interactions between a Nanomagnet and a Photonic Cavity} published in a 2010 PRL \cite{Soykal2010}, Michael Flatt\'{e} and his PhD student \"{O}ney Orhunç Soykal at the University of Iowa analyze the interaction of a nanomagnet with a single photonic mode of a cavity in a fully quantum-mechanical treatment. The result is very interesting since it predicts an exceptionally large quantum-coherent magnon-photon coupling which reaches the strong coupling regime (which means the coupling strength exceeds the dissipation of the coupled system). Appealing to everyone interested in quantum physics, they showed that the magnon-photon coupling leads to eigenstates that correspond to entangled states of spin orientation and photon number. In such mixed quantum states of light and matter, Rabi oscillations occur in which initial (coherent) states of definite spin and photon number evolve dynamically to produce large oscillations in the microwave power.

Three years later in 2013, the Munich group of Hans Huebl and Sebastian Goennenwein at the Walther-Meißner-Institut published in PRL the first experimental result on magnon-photon coupling \cite{Huebl2013}. In their paper entitled \textit{High Cooperativity in Coupled Microwave Resonator Ferrimagnetic Insulator Hybrids}, Huebl et al. demonstrate how to use microwave transmission experiment to measure at 50 mK the strong coupling between magnons in a yttrium iron garnet (YIG) and photons in a superconducting coplanar microwave resonator made from Nb. 

Huebl is an expert in quantum microwave devices, and Goennenwein is a driving force of spin mechanics. Other Viking explores arrive the Newfoundland of Cavity Spintronics from very different galaxies of physics.

In the galaxy of superconducting device and quantum computing, Yasunobu Nakamura at the University of Tokyo has been a star since he published the paper \textit{Coherent control of macroscopic quantum states in a single-Cooper-pair box} in 1999 in Nature \cite{Nakamura1999}. In that work, Nakamura \textit{et al}. measured the Rabi oscillation of a superconducting device. That was the first demonstration of a practical solid-state qubit which employed the macroscopic quantum coherence of the two charge states of a \textit{single-Cooper-pair box}. Such a superconducting device was what Schoelkopf later used in his solid-state QED experiment \cite{Wallraff2004}. In another galaxy, Schoelkopf's young colleague Hong-Xing Tang at Yale University is a rising star of nano-electromechanical systems and quantum optics, whose group has created integrated photonic and nanomechanical devices for quantum force measurements and quantum limited displacement sensing. Via the wormhole of quantum coherence, both Nakamura and Tang entered the field of Cavity Spintronics, setting their feet on the soil of the magnons.

Both of their studies of magnon-photon coupling were published in 2014 in PRL, showing how to tune the coupling strength. In Nakamura's paper entitled \textit{Hybridizing Ferromagnetic Magnons and Microwave Photons in the Quantum Limit} \cite{Tabuchi2014}, the Tokyo team used a variable-sample method. Their experiment was performed at cryogenic temperatures in the quantum regime where the average number of thermally or externally excited magnons and photons was less than one.  By changing the volume of the ferromagnetic sample set in the cavity, they confirmed that the coupling strength was proportional to the square root of the number of spins in the cavity. That's the smoking gun of quantum strong coupling as predicted by the quantum theory. The experiment of Tang's group, in contrast, was performed at room temperature. In the paper \textit{Strongly Coupled Magnons and Cavity Microwave Photons} \cite{Zhang2014}, the Yale team used a variable cavity method to show that, by scaling down the cavity size, the coupling strength can be tuned to reach an ultrahigh cooperativity (characterizing the ratio of coupling over all damping) of 12 600. They also showed other interesting dynamic features such as classical Rabi-like oscillations, magnetically induced transparency, and the Purcell effect. Alerted readers may notice that Tang \textit{et al}. carefully used the term \textit{classical Rabi-like oscillations} \cite{Zhang2014} instead of Rabi oscillations which usually refers to quantum coherence. Readers of Nakamura's paper may find a note added to their paper \cite{Tabuchi2014}, in which the distinction between the quantum and classical regime was discussed. That's the point to which we will be back in the next section.	
 
All these pioneering works done in Germany, Japan, and the USA were performed by using either the microwave transmission or reflection spectroscopy. The vision of Silsbee has taught us that the highway of spintronics is spin transport \cite{Johnson1985}. Building this highway for Cavity Spintronics requires developing an electrical method to detect the magnons coupled with photons. Here come the Canadian settlers.

For decades, electrical detection of charge dynamics has been extensively used for studying semiconductor physics \cite{Holland2004}. Since 2004, we have set out to expand this technique to study spin and magnetization dynamics by developing microwave photo-conductivity, photo-current, and photo-voltage spectroscopies \cite{Gui2005,Gui2007}. By then, it was nearly a no man's land in magnetism, but through a decade of effort by many spintronics groups worldwide, this branch of magnetism is now booming with diverse methods available \cite{Bai2014}. In the paper entitled \textit{Spin Pumping in Electrodynamically Coupled Magnon-Photon Systems} \cite{Bai2015}, which was published in 2015 as a PRL Editor's Suggestion, Lihui Bai and Michael Harder at the University of Manitoba pick up the tool of spin pumping to study the magnon-photon coupling. This is achieved by designing a special microwave cavity as schematically shown in Fig. 1(a). This set-up enables both microwave transmission measurement of the cavity, as well as microwave photo-voltage measurement of the magnet. Setting in the cavity a bilayer device of YIG/Pt fabricated by the group of John Xiao at the University of Delaware, the microwave photo-voltage measured in Pt probes the spin current generated by the FMR in YIG. This enables studying the impact of magnon-photon coupling on the spin transport. In an insightful viewpoint article entitled \textit{Electrical signal picks up a magnet's heartbeat} \cite{Huebl2015}, Huebl and Goennenwein predict that such an electrical detection method could lead to new ways of reading out the quantum state of a magnet with compact electronics.

\begin{figure}[t]
\includegraphics[width = 8.5 cm]{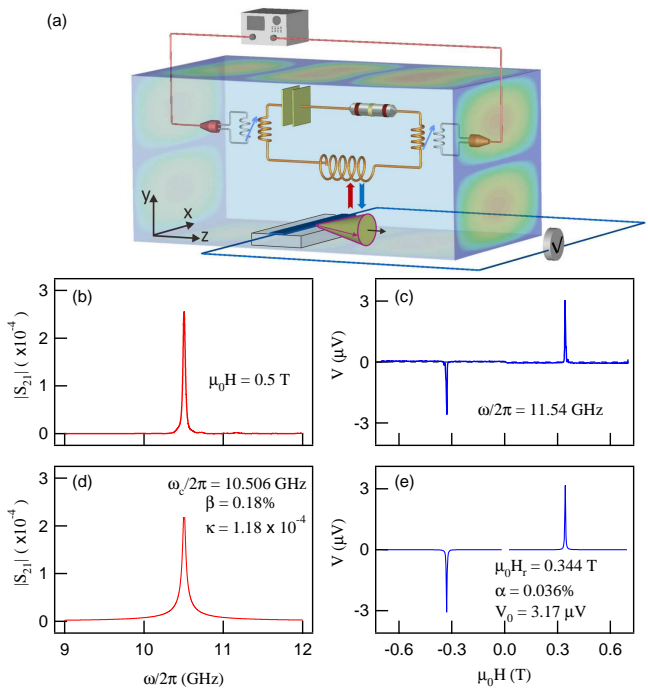}
\caption{\textbf{Electrical detection of the Cavity Magnon Polariton (adopted from Ref. \cite{Bai2015})}. (a) Sketch of the experimental setup and theoretical model for studying the cavity magnon polariton (CMP). The artificial LCR circuit models the microwave current carried by the cavity mode, which couples with the magnons in the magnetic material. Such a CMP can be detected by measuring (b) microwave transmission spectrum S21, and (c) photo-voltage V. (d) S21 and (e) V calculated theoretically.}
\label{sketch}
\end{figure}

While this is a very exciting future perspective, the experiment of Bai \textit{et al}. was performed in the classical regime at room temperature. Not only that, in contrast to previous studies guided by the theory of quantum strong coupling \cite{Soykal2010}, the Manitoba-Delaware collaboration did not follow the quantum map, instead, a concise theory was developed showing that the magnon-photon coupling in general can be described on the classical footing of electrodynamics \cite{Bai2015}. This gives Cavity Spintronics an alternative classical map.

\section{Quantum or Classical, This is the Question}

Understanding the transition from classical to quantum physics has always been a fascinating topic [10,11]. In the early days of quantum mechanics, it was thought that there was a fundamental difference between the microscopic systems that obey quantum mechanics and the macroscopic world of classical physics. Discoveries of macroscopic quantum states have changed that simple view. To verify macroscopic quantum systems, Rabi oscillations are often measured, in which the quantum superposition of distinct quantum states is probed via the coherent oscillation between them. Since the superposition of states is a purely quantum effect that has no classical analogue, Rabi oscillations enable the fundamental demonstration of quantum systems. Nakamura's experiment in 1999 was a tour de force example \cite{Nakamura1999}.

Initial interest in Cavity Spintronics was based on such a quantum perspective \cite{Soykal2010}. Via the quantum strong coupling of magnons and photons which generates entangled states of spin orientation and photon number, quantum information can be easily transferred between the light and the magnet via Rabi oscillations. Such quantum entangled states require cryogenic conditions, because the interaction of magnons with phonons through spin-orbit coupling limits the dephasing time at room temperature \cite{Soykal2010}. That was one of the reasons that the magnon-photon coupling experiments were initially performed at extremely low temperatures.

Tang \textit{et al}.'s experimental result came as a big surprise. Not only did they demonstrated that the ultrastrong coupling regime can be reached at room temperature, but they also observed directly in the time domain up to 10 cycles of microwave oscillations induced by the magnon-photon coupling. Something must be happening beyond the scope of the quantum map. Alerted by this, Tang \textit{et al}. carefully coined the term \textit{classical Rabi-like oscillations} \cite{Zhang2014}, and Nakamura \textit{et al}. added a similar note \cite{Tabuchi2014}. But the specific origin of classical magnon-photon coupling was not clear.

That mystery was solved by the Manitoba-Delaware collaboration \cite{Bai2015}, which singled out the classical origin of magnon-photon coupling: the phase correlation of classical electrodynamics. Quantum coherence stems from entangled states evolving according to Schrodinger's equation, but Maxwell's equations of macroscopic electromagnetic fields also contain classical coherence. In the wave physics community, the close resemblance between Schrodinger's equation and Maxwell's equations has led to the development of photonic crystals, which uses the analogy of coherent effects between the classical wave of light and the quantum wave of electrons. The same resemblance implies that the magnon-photon coupling may be based either on the quantum coherence of entangled states or the classical coherence of macroscopic fields.

Fig. 1 schematically shows the general picture of the classical coupling: without coupling, the magnetization precession $m(t)$ in the magnet is determined by the LLG equation \cite{Bai2015}, and the cavity resonance is modeled by an artificial $LCR$ circuit that carries the microwave current $j(t)$. With magnon-photon coupling, $m(t)$ induces a voltage in the $LCR$ circuit. That's the Faraday's law. On the other hand according to Ampère's law, $j(t)$ produces a magnetic field $h(t)$ that places a torque on the magnetization. Such a classical mutual coupling can be explicitly modelled as a concise 2$\times$2 eigenvalue equations \cite{Bai2015}, which accurately reveals the key physics of phase correlation between $m(t)$ and $j(t)$ governed by the electrodynamics. This general picture is independent of the sample or cavity details. It provides a classical map of the Cavity Spintronics. In consistent with such a general picture \cite{Bai2015}, an 1D scattering theory was independently developed for studying the classical coupling in the special case of 1D cavity geometry \cite{Cao2015}.

With the classical map, it is clear that the magnon-photon coupling measured at room temperature is a classical electrodynamic effect. A new intriguing question arises of how to properly distinguish the \textit{classical Rabi-like oscillations} of macroscopic electrodynamic fields from quantum Rabi oscillations of entangled states. Observing Rabi oscillations in the coupled magnon-photon system, even at cryogenic temperatures with few photons, is no longer an irrefutable proof of quantum coherence of entangled states. New experiments need to be designed to search for exclusive quantum features, and progress is been making \cite{Tabuchi2015}. In modern physics, probing the quantum superposition principle at the borderline to classical physics has been a driving force \cite{Wineland2013,Haroche2013,Ketterle2002}. Now, Cavity Spintronics adds new fuel to this powerful engine.

As intriguing as the quantum question, the classical map reveals a new wormhole. As shown in the classical model \cite{Bai2015}, the eigen vector of the coupled magnon-photon mode is a linear combination of the rf magnetic field $h(t)$ and rf magnetization $m(t)$. This is by definition the magnon polariton \cite{Mills1996}. Such a new insight links Cavity Spintronics with the physics of Cavity Polariton which is an exciting frontier of semiconductor research.

\section{The Renascence of Cavity Polaritons}

A polariton is an optical effect arising when light couples to a material that has a macroscopic polarization or magnetization. This concept was developed by Kun Huang. In 1951, the 32-year old Chinese physicist published two seminal papers: \textit{Lattice Vibrations and Optical Waves in Ionic Crystals} \cite{Huang1951a} and \textit{On the Interaction between the Radiation Field and Ionic Crystals} \cite{Huang1951b}, in which he introduced the term \textit{polariton} to describe the classical coupling between electromagnetic wave and macroscopic polarization. With this concept Huang derived the infrared optical properties of lattice vibration in ionic crystals. This concept is so fundamental that it soon became the 'basic knowledge' of solid-state textbooks. It ubiquitously explains the interaction between electromagnetic wave and elementary excitations in materials such as phonons, excitons, and magnons \cite{Mills1996}.

A renascence placed polariton at the frontier of semiconductor research. That started in 1992 when Claude Weisbuch \textit{et al}. published in PRL an article entitled \textit{Observation of the Coupled Exciton-Photon Mode Splitting in a Semiconductor Quantum Microcavity} \cite{Weisbuch1992}. It showed that the strong coupling of the exciton and cavity photon in a semiconductor microcavity leads to the formation of a cavity exciton polariton, a half-light, half-matter bosonic quasi-particle. Since then, research in that frontier has led to remarkable breakthroughs in both basic and applied research. For examples, Bose-Einstein condensation \cite{Kasprzak2006} and superfluidity \cite{Amo2009} of cavity exciton polaritons are discovered at standard cryogenic temperatures, parametric scattering driven by a polariton condensate is found at the room-temperature \cite{Xie2012}, and the electrically pumped cavity polariton laser is developed showing unsurpassed properties due to the physics of coherent strong coupling \cite{Schneider2013}.

So far, experiments on cavity polaritons are mainly performed in semiconductor materials at optical frequencies, using the strong coupling of excitons with photons \cite{Weisbuch1992,Kasprzak2006,Amo2009,Xie2012,Schneider2013}. Now, the classical map of Cavity Spintronics reveals a new type of cavity polariton: the \textit{Cavity Magnon Polariton} \cite{Bai2015} that is based on magnetic materials operating at microwave frequencies. Soon, the strong analogies between the polariton physics of the two different systems may merge the studies of Cavity Exciton Polaritons with the growing interest in Cavity Magnon Polaritons. This may not only lead to a better understanding of the fundamental physics of strong coupling between magnons and photons, but along such an adventure in basic research, new microwave and spintronic applications \cite{Kaur2015} that are beyond our imagination will emerge.

\section{Closing Remarks and Future Perspectives}

In summary, advances in magnetism, nanotechnology, and light-matter interaction have created a new frontier of condensed matter research studying Cavity Spintronics. Via the quantum physics of spin-photon entanglement on the one hand, and via the classical electrodynamic coupling on the other, this frontier merges the progress in spintronics with the advances in cavity QED and cavity polaritons. This article is focused on the historical context of this frontier by tracing it back to some of the most courageous, inspiring, and seminal work in the history of spintronics, cavity QED and polaritons.

Looking forward from the Canadian perspective, the development of Cavity Spintronics may benefit from the remarkable progress in Cavity Optomechanics, made by Canadian physicists in the groups of Mark Freeman and John Davis at the University of Alberta, Paul Barclay at the University of Calgary, and Aashish Clerk at McGill University. Future study of quantum physics in Cavity Spintronics may benefit from using unique quantum magnetic materials, grown by Chris Wiebe at the University of Winnipeg and the members of the Brockhous Institute for Materials Research at McMaster University. Last but not least, future development of on-chip Cavity Spintronics will benefit from the research on miniaturization of microwave circuits, in which Canadian engineers, such as Lot Shafai and Greg Bridges at the University of Manitoba, are making remarkable contributions. With the wonderful twilight appearing in the dawn sky of Cavity Spintronics, perhaps an enjoyable way of imagining its future is to listening to the song \cite{Celtic2008}:\\

\textit{I am the dawn, I'm the new day begun}

\textit{I bring you the morning, I bring you the sun}

\textit{I hold back the night and I open the skies}

\textit{I give light to the world, I give sight to your eyes}

\textit{I am the sky and the dawn and the sun}

\section{Acknowledgements}

I would like to thank my co-workers L.H. Bai, M. Harder, B.M. Yao, S. Kaur, and Y.S. Gui for their contributions. I also thank my friends and colleagues M. Johnson, J. Sirker, S. Goennenwein, H. Huebl, Y. Nakamura, H.X. Tang, T. Dietl, M. Freeman, H. Guo, J. Page, C. Sch\"{u}ller, Y. Xiao, J.Q. Xiao, W. Lu, and Z.H. Chen for discussions. Financial support from NSERC, CFI and NSFC's oversea programm are gratefully acknowledged.

%%%%%%%%%%%%%%%%%%%%%%%%%%%%%%%%%%%%%

\end{document}